\documentclass[conference,anonymous]{IEEEtran}
\IEEEoverridecommandlockouts

\usepackage{hyperref}
\hypersetup{
    colorlinks=true,
    linkcolor=black,
    filecolor=magenta,
    citecolor=black,
    urlcolor=blue,
    }

\usepackage{graphicx}
\usepackage{balance}
\usepackage{algorithmic}
\usepackage{booktabs}
\usepackage{xcolor}
\usepackage{colortbl}
\usepackage{makecell}
\usepackage{wrapfig}
\usepackage{caption}
\usepackage{url}
\usepackage{amsmath}
\usepackage{orcidlink}

\newcommand{\approach}[0]{BERTiMuS}

\begin{document}

\title{Simulink Mutation Testing using CodeBERT}

\author{\IEEEauthorblockN{Jingfan Zhang, Delaram Ghobari,  Mehrdad Sabetzadeh\orcidlink{0000-0002-4711-8319}, Shiva Nejati\orcidlink{0000-0002-0281-8231}}
\IEEEauthorblockA{School of Electrical Engineering and Computer Science\\
University of Ottawa, Canada\\
\{jzhan665,dghob088,m.sabetzadeh,snejati\}@uottawa.ca}
}

\maketitle
\begin{abstract}
We present \approach, an approach that uses CodeBERT to generate mutants for Simulink models. \approach\ converts Simulink models into textual representations, masks tokens from the derived text, and uses CodeBERT to predict the masked tokens. Simulink mutants are obtained by replacing the masked tokens with  predictions from CodeBERT. We evaluate \approach\ using Simulink models from an industrial benchmark, and compare it with FIM -- a state-of-the-art mutation tool for Simulink. We show that, relying exclusively on  CodeBERT, \approach\  can generate the block-based Simulink mutation patterns documented in the literature. Further, our results indicate that: (a)~\approach\ is complementary to FIM, and (b) when one considers a requirements-aware notion of mutation testing, \approach\ outperforms FIM. 
\end{abstract}


\section{Introduction}
\label{sec:intro}
Large language models (LLMs) are increasingly being used to automate and enhance software testing activities, including mutation testing~\cite{papadakis2019mutation}. LLMs have been used to generate code mutants through two main methods: Masked Language Modeling (MLM)~\cite{degiovanni2022mubert,khanfir2023efficient} and prompt-based techniques~\cite{dakhel2024effective,brownlee2023enhancing}.  MLM-based mutation generation uses models such as CodeBERT~\cite{feng2020codebert} to mask tokens within a code sequence and predict alternatives that best fit the context. On the other hand, prompt-based approaches create specific prompts for conversational, generative LLMs such as GPT-4~\cite{achiam2023gpt} to generate code mutants. The hypothesis is that  LLMs can generate mutants that conform to programmers' implicit rules and conventions, potentially outperforming syntax-based fault-seeding approaches to mutation testing.

LLMs have shown to be effective for analyzing text-based programming languages. Nevertheless, their effectiveness with graphical languages, e.g., Simulink~\cite{simulink}, has not been  studied much. Simulink is widely used in the cyber-physical systems (CPS) domain for modeling and simulation. There are already mutation tools for Simulink models, e.g., SIMULATE~\cite{PillRWN16} and the fault injection and mutation tool (FIM)~\cite{bartocci2022fim}. These tools offer mutation operators based on known Simulink fault patterns, as well as mechanisms for defining new fault patterns. However, unlike code, which benefits from open-source repositories for identifying real-world fault patterns~\cite{just2014defects4j}, Simulink models lack extensive public repositories, hindering the identification of real-world faults. Most Simulink fault types are derived from code mutation operators or specialized mathematical mutation operators.  Given the success of LLMs in  enhancing code mutation testing~\cite{degiovanni2022mubert,khanfir2023efficient,dakhel2024effective,brownlee2023enhancing},  it remains to be seen if and how LLMs can also improve mutant generation for Simulink models, \hbox{especially with limited fault-pattern resources in this domain.}

In this paper, we propose \approach: a \textbf{BERT}-ass\textbf{i}sted approach to \textbf{Mu}tant generation for \textbf{S}imulink. \approach\ employs CodeBERT, an LLM specialized in interpreting and generating  code, to generate mutants for Simulink models.  Similar to the recent approach of converting graphical structures into text~\cite{GrieblFOFJ23,LuitelNS24}, \approach\ converts graphical Simulink models into textual representations that are amenable to processing by LLMs. \approach\ generates mutants using an MLM-based strategy by masking and replacing tokens based on CodeBERT's predictions. To ensure that CodeBERT is familiar with the structure of textual representations obtained from Simulink models, we  fine-tune the MLM by further training the pre-trained CodeBERT model on a large corpus of text derived from third-party Simulink models available at SLNET~\cite{ShresthaCC22} before using it for mutant generation.

An important goal of mutation testing is to generate mutants that align with the programmer's intent or the requirements of the system under test~\cite{papadakis2019mutation}. Bartocci et al.~\cite{bartocci2023property} propose a new notion of mutant killing, asserting that a test case kills a mutant if executing the test case on the mutant leads to a requirement violation, whereas executing the same test case on the original code does not result in such a violation. This contrasts with the classical notion of mutant killing, where a mutant is considered killed by a test case if the outputs from executing the test case on the mutant and on the original code differ. Since requirements are often available for Simulink models,  we evaluate \approach\ for  both \emph{classical} and \emph{requirements-aware} notions of mutant killing.
\emph{A mutation  is requirements-aware if it causes a requirement violation.}

We evaluate \approach\ using five Simulink models from industrial benchmarks~\cite{9218211,bartocci2022fim} and compare it with the state-of-the-art fault injection and mutation (FIM) tool for Simulink~\cite{bartocci2022fim}. We show that \approach\ covers all known block-based Simulink mutation patterns documented in the literature~\cite{rajan2008requirements,le2014mutation,binh2012mutation,he2011test,papadakis2019mutation,matinnejad2018test}. Furthermore, we demonstrate that \approach\ and FIM are complementary because each generates mutants that lead to the selection of different sets of test cases. Specifically, the test cases selected using mutants from \approach\ do not entirely overlap with those selected using mutants from FIM, and vice versa. Finally, under the requirements-aware notion of mutation testing, \approach\ significantly outperforms FIM. Specifically,  under this notion, test cases selected using FIM mutants failed to kill twice as many \approach\ mutants as the number of FIM mutants not killed by test cases selected using \approach\ mutants. This indicates that \approach\ generates a larger number of mutants representing requirements-relevant faults that are harder to detect than those generated by FIM.

\section{BERT-assisted Simulink Mutation}
\label{sec:app}
\begin{figure}[t]
	\centering
\includegraphics[width=0.85\columnwidth]{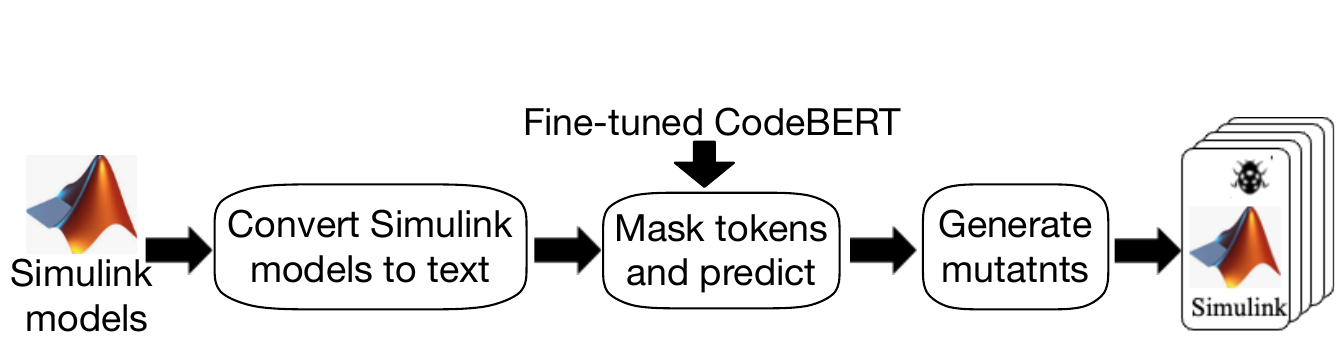}		
\vspace*{-.2cm}
	\caption{Overview of \approach}\label{fig:framework}

\end{figure}

\begin{figure}
  \centering
    \includegraphics[width=0.85\columnwidth]{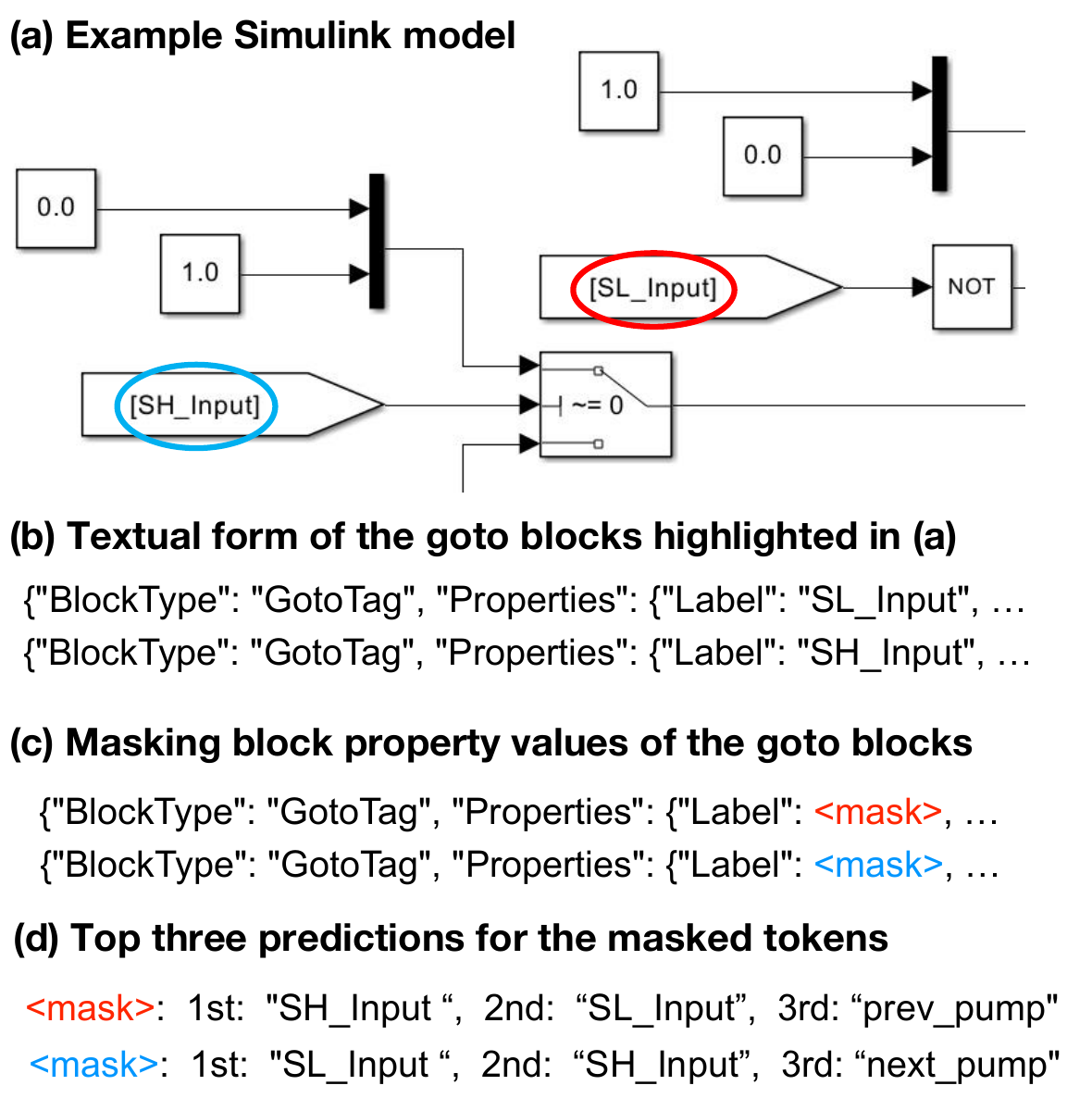} 
    \vspace*{-.1cm}
  \caption{Illustration of token masking by \approach}
  \label{fig:simseq}
  \vspace*{-.4cm}
\end{figure}

Figure~\ref{fig:framework} shows an overview of \approach. The core of \approach\ is CodeBERT fine-tuned on a large corpus of  textual representations of Simulink models. \approach\ consists of the following three steps:

\textbf{Step~1. Convert to textual format.} \approach\ converts the input Simulink model, which is graphical, into a textual format. Simulink models already have an XML-based textual representation supported by MATLAB/Simulink. The underlying XML schema envisages different attributes, covering the semantics, syntax, and graphical information, such as block name, type, ID, block properties, and block coordinates for visual rendering. \approach{} creates a simplified representation based on this content, retaining only the context needed for CodeBERT to mutate the Simulink blocks. \approach{} stores this simplified representation in JSON format.
Figure~\ref{fig:simseq}(b) shows the textual representations that \approach\ generates corresponding to the two GotoTag blocks encircled in red and blue in Figure~\ref{fig:simseq}(a). In these textual representations, the GotoTag labels, \texttt{SL\_Input} and \texttt{SH\_Input},  are stored in nested objects under the \texttt{Properties} key.

\textbf{Step~2. Mutate by masking and predicting.}   The textual representations created in Step~1 have a list of properties for each block that capture the semantic elements of that block, e.g., the labels of the GotoTag blocks in  Figure~\ref{fig:simseq}(b). To build Simulink mutants, \approach\ masks each block property and uses CodeBERT to predict alternatives for the masked property.  For example, Figure~\ref{fig:simseq}(c) shows the masking of the labels  of the GotoTag blocks encircled in Figure~\ref{fig:simseq}(a). Since CodeBERT is based on BERT, a bidirectional transformer model, it can capture the semantic relationships between masked tokens and their surrounding tokens.  \approach{} uses CodeBERT, after being fine-tuned on a large corpus of Simulink textual representations, to predict the masked tokens in Figure~\ref{fig:simseq}(c). The text before and after each masked token, e.g. the names, types and properties of other blocks, provides context that guides CodeBERT's predictions. The top predictions by CodeBERT for the two masked tokens in Figure~\ref{fig:simseq}(c) are shown in Figure~\ref{fig:simseq}(d). Note that \texttt{next\_pump} and \texttt{prev\_pump} are the labels  of some other blocks in the original Simulink model that are not shown in Figure~\ref{fig:simseq}(a).

\textbf{Step~3. Generate mutants.} \approach\ generates mutants by replacing the masked token with each of the top $k$ CodeBERT predictions that differ from the original value, discarding any mutants that do not compile. For the example in Figure~\ref{fig:simseq}(d), after discarding the predictions that match the original values of the masked tokens, \approach{} generates mutants where \texttt{SL\_Input} is replaced by \texttt{SH\_Input} and \texttt{prev\_pump}, and \texttt{SH\_Input} is replaced by \texttt{SL\_Input} and \texttt{next\_pump}.  These mutants are compilable and introduce logical errors. Furthermore,  they are \emph{requirements-aware} since they lead to the violation of some requirement of the original model in Figure~\ref{fig:simseq}(a). Specifically, the variables \texttt{SL\_Input} and \texttt{SH\_Input} represent the low and high safety levels, respectively, and the violated requirements pertain to the high and low safety levels of the liquid pumped into a tank controlled by the Simulink model in Figure~\ref{fig:simseq}(a). 
Although these mutants can also be generated programmatically, doing so requires significant implementation effort and involves several case-by-case hard-coded decisions. \approach{} can generate them using a simple mask-and-predict strategy.

\section{Evaluation Setup}
\label{sec:eval}

\textbf{RQ1} \emph{How do the mutants generated by \approach{} compare with those generated by existing mutation patterns for Simulink models?} For this research question, we compare the mutants generated by \approach\ with those generated by existing  mutation patterns for Simulink~\cite{rajan2008requirements,le2014mutation,binh2012mutation,he2011test,papadakis2019mutation,matinnejad2018test}.

\textbf{RQ2.} \emph{Can \approach{} generate more effective mutants compared to the state-of-the-art mutant generation approach for Simulink?} We compare \approach\ with the fault injection and mutation (FIM) tool~\cite{bartocci2022fim}.

To compare \approach\ with FIM, we consider the test cases selected from a reference test suite for killing mutants produced by each approach. We then assess (1) whether the test cases selected by one approach subsume those generated by the other or whether these approaches find complementary test cases, and (2) whether the test cases selected by each approach can kill the mutants generated by the other approach.

As discussed in Section~\ref{sec:intro}, Simulink mutants may be killed in two ways: \emph{classical} (also known as  \emph{output-based}) mutation testing, where a mutant is killed by a test case if it produces outputs different from the original model for that test case, and \emph{requirements-aware} mutation testing, where a mutant is killed by a test case if the test case reveals that the mutant violates a requirement satisfied by the original model~\cite{bartocci2023property}. The requirements-aware notion is stronger than the classical one: If a test case kills a mutant under the requirements-aware notion, it also does so under the classical notion, but not vice versa. Furthermore, the requirements-aware notion determines whether a mutant is relevant to a requirement and whether a test case that kills a mutant is indeed revealing a realistic fault. In the absence of real-world fault repositories for Simulink models, for RQ2, we consider requirements-aware mutation testing alongside classical mutation testing to more effectively determine the relevance and usefulness of the mutants.

\textbf{Study subjects.} Table~\ref{tab:simulinkmodels} presents the characteristics of our Simulink subject models, including the number of blocks and the number of requirements used to assess these models in requirements-aware mutation testing. The first three models are  from the Lockheed Martin benchmark~\cite{9218211}, and the remaining two are provided with our baseline tool~\cite{bartocci2022fim}.

\begin{table}[t]
\captionsetup{skip=0.1cm}
\caption{Features of our Simulink subject models.}~\label{tab:simulinkmodels}
\vspace*{-.4cm}
\begin{center}
\scalebox{0.69}{\begin{tabular}{|l|p{8.4cm}|l|l|}
\hline
\textbf{Name} & \textbf{Description} & \textbf{\#blocks} & \textbf{\#Reqs} \\ \hline
 Tustin                             & A numeric model that computes integral over time~\cite{9218211}.  & 57 & 5  \\ \hline

 Twotanks    & A controller regulating the incoming and outgoing flows of two tanks~\cite{9218211}.  & 498 & 32   \\ \hline 
  FSM    & Controls the autopilot mode in case of some environment hazard~\cite{9218211}.
  & 303  &13  \\ \hline
  
    ATCS    & An automatic transmission controller system~\cite{bartocci2022fim}.
  & 65 &1  \\ \hline
  
      AECS    & An aircraft elevator control system~\cite{bartocci2022fim}.
  & 825 & 2   \\ \hline
\end{tabular}}
\end{center}
\vspace*{-.3cm}
\end{table}

\textbf{Fine-tuning CodeBERT on Simulink.} Although CodeBERT is pre-trained on various programming languages and can interpret code semantics well, it has not been specifically trained on Simulink  models and, hence, may lack knowledge pertinent to Simulink. To enhance CodeBERT's performance on Simulink, we fine-tuned CodeBERT on a large corpus of text obtained  from 2,611 Simulink models in the SLNET repository~\cite{ShresthaCC22}. To obtain this corpus, we applied the same conversion as in Step~1 of \approach\ (Figure~\ref{fig:simseq}) to the models in SLNET. Following common practices in code-representation learning~\cite{feng2020codebert}, we use the masked language modeling method~\cite{devlin2018bert} for 15 epochs. During fine-tuning, we randomly masked 15\% of the tokens in the textual representation of Simulink models and tasked CodeBERT with predicting the masked tokens.

\section{Experiments and Results}
\textbf{RQ1 experiments and  results.} We use \approach\ to generate mutants for the models in Table~\ref{tab:simulinkmodels}. For each masked value, we consider the top three predictions that are not identical to the original masked value. On average, $80$\% of the generated mutants were compilable. After removing non-compilable mutants, we obtained a total of 387 compilable mutants for our subject models.

For RQ1, we consider block-based mutation patterns, i.e., those related to modifying individual blocks. \approach\ is not designed to perform link-based mutations, e.g., changing links between blocks or swapping two blocks. Such mutations can be effectively handled by existing rule-based mutation operators. Table~\ref{tab:rq1} summarizes the main block-based mutation patterns for Simulink that we have collected by extensively surveying the literature~\cite{rajan2008requirements,le2014mutation,binh2012mutation,he2011test,papadakis2019mutation,matinnejad2018test}.

\begin{table}[t]
\captionsetup{skip=0.1cm}
\caption{Block-based Simulink mutation patterns.}~\label{tab:rq1}
\scalebox{0.63}{\begin{tabular}{|p{6.7cm}|p{6.5cm}|}
\hline
 Mutate signal data types~\cite{matinnejad2018test} &   Mutate GoTo/From blocks~\cite{matinnejad2018test}  \\ \hline 

 Mutate ``Saturate on integer overflow"~\cite{matinnejad2018test}  
& 
 Mutate constant and gain values~\cite{matinnejad2018test, jia2010analysis, he2011test, le2014mutation,binh2012mutation} \\ \hline
 Mutate  math, relational and logical operator blocks\cite{matinnejad2018test,jia2010analysis,he2011test,le2014mutation,binh2012mutation} & 
 Mutate initial conditions and sample time~\cite{matinnejad2018test,binh2012mutation}  \\ \hline
  Mutate Stateflow transition  conditions~\cite{matinnejad2018test,jia2010analysis,he2011test,le2014mutation}  &   Mutate Stateflow variable names~\cite{matinnejad2018test,le2014mutation,binh2012mutation} \\ \hline
    Mutate Stateflow actions~\cite{matinnejad2018test,le2014mutation,binh2012mutation,jia2010analysis} &  Mutate Stateflow keywords~\cite{matinnejad2018test,binh2012mutation} \\ \hline  
\end{tabular}}
\vspace*{-.4cm}
\end{table}

To address RQ1, we examined whether the 387 mutants generated by \approach\ could be produced using the mutation patterns in Table~\ref{tab:rq1}. Our analysis confirms that all these mutants can be generated using the 10 patterns listed in Table~\ref{tab:rq1}. While \approach\ does not generate any mutants beyond the patterns in Table~\ref{tab:rq1}, our investigation shows that \approach, due to its fine-tuned knowledge of Simulink structure and its access to context at the time of masking, generates mutant instances that are unlikely to be generated based on syntactic and manual mutation rules. Examples of such mutant instances illustrated in Figure~\ref{fig:simseq} where  \approach\ replaces a masked label or variable name in a block or structure with a label or variable name already used elsewhere in the model such that the new label or variable has a consistent type with that of the masked element. 
As another example, \approach\ could alter masked keywords in Stateflows (Simulink state machines) into consistent counterparts, e.g., changing \texttt{before} to \texttt{after} or \texttt{at}, yielding compilable mutants  \hbox{with realistic logical flaws.}

\textbf{RQ2 experiments.} To answer RQ2, we compare \approach\  and our baseline, the FIM tool, with respect to the set of test cases that can be generated using the mutants produced by these approaches. We use  FIM to generate mutants for the models listed in Table~\ref{tab:simulinkmodels}.  In total, FIM produced 594 compilable mutants for our subject models in Table~\ref{tab:simulinkmodels}.  Since FIM generates more mutants than \approach, i.e., 387 compilable mutants by \approach\ versus 594 compilable mutants by FIM, it would also generate more test cases, as more test cases are required to kill its mutants.

To objectively compare FIM and \approach, we use an experimental procedure previously proposed to compare classical mutation testing with CodeBERT-based mutation testing~\cite{degiovanni2022mubert}. Similar to our work,  in code mutation testing, baselines that use  syntax-based operators generate more mutants than CodeBERT-based mutation generators~\cite{degiovanni2022mubert}. This procedure uses mutants from each approach to select a subset of  test cases from a reference test suite for each subject Simulink model, thus enabling a comparison of the selected test cases by each approach.  The procedure begins with a reference test suite $\mathit{TS}$ for each Simulink model. For both  FIM and \approach, we select a minimal subset $T \subseteq \mathit{TS}$ of test cases that can kill as many mutants generated by each approach as possible. We then compare the overlap between the minimal test subsets $T$ generated by the two approaches to determine whether one approach subsumes the other. We repeat this procedure for both notions of mutant killing, i.e., classical and requirements-aware. To reduce the impact of random selection of mutants, we repeat this procedure five times for each mutant generation method, each Simulink model, and each notion of mutation testing.   We generate the reference test suites using adaptive random testing, as used in earlier studies~\cite{bartocci2023property,matinnejad2018test,NejatiGMBFW19,MatinnejadNBBP13}.

\begin{figure}[t]
  \centering
  \includegraphics[width=0.95\linewidth]{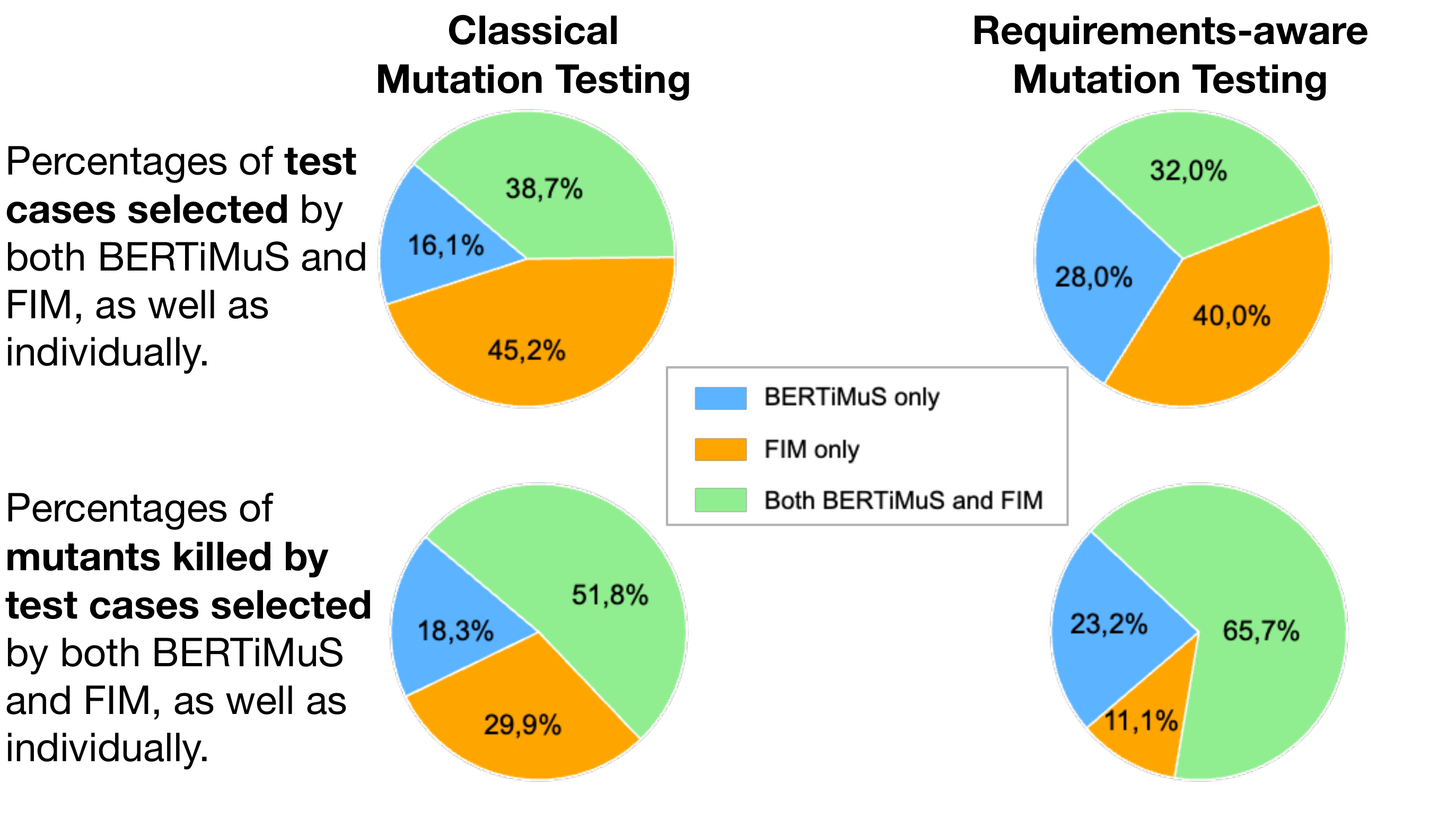}
\caption{Comparing  \approach\ and FIM based on the average number of test cases selected by each, and the average number of mutants  killed by those test cases. }\label{fig:test_cases_overlap}
\vspace*{-.2cm}
\end{figure}

 \textbf{RQ2 results.} The pie charts in the \emph{first row} of Figure~\ref{fig:test_cases_overlap}  show the average percentages of test cases selected by \approach\ mutants only, by FIM mutants only, and by both. The pie charts in the \emph{second row} of Figure~\ref{fig:test_cases_overlap} show the average percentages of \approach\ mutants killed by test cases selected by \approach\ but not by those selected by FIM, highlighted in blue; the average percentages of FIM mutants killed by test cases selected by FIM but not by those selected by \approach, highlighted in orange; and the average percentages of mutants from both approaches that could be killed by test cases selected by both approaches, highlighted in green.
The two charts on the left correspond to the classical (output-based) notion of mutant killing, and the two on the right correspond to the requirements-aware notion of mutant killing. In addition, Table~\ref{tab:five models} shows the mutation scores, i.e., the percentages of killed mutants over all the killable mutants, for \approach\ and FIM according to classical and requirements-aware mutation testing, as well as the total number of test cases selected from the reference test suites to kill mutants. The table also shows the actual number of killed mutants by each approach that could not be killed by the test cases selected by the other approach. The averages in Table~\ref{tab:five models} are computed over five runs; however, we note that there were almost no differences across the five runs in terms of the number of test cases selected from the reference sets, and subsequently, the number of killed mutants by these test cases.  Hence, the averages are presented as integers.

\begin{table}[t]
\captionsetup{skip=0.1cm}
\caption{Average number of selected test cases and mutation scores for  \approach\ and FIM  for the models in Table~\ref{tab:simulinkmodels}}\label{tab:five models}
\scalebox{.68}{\begin{tabular}{|p{3.3cm}|c|c|c|c|} 
\hline
{} & \multicolumn{2}{c|}{BERTiMuS} & \multicolumn{2}{c|}{FIM} \\ \cline{2-5} 
                           & classic  & req-aware & classic   & req-aware  \\ \hline
\# of killable mutants  & 311 & 311 & 402 & 402 \\ \hline 
Average \# of selected test cases  from the reference sets        & 19  & 9   & 38  & 14    \\ \hline

Average \# of killed mutants by the 
selected test cases    & 282  & 239 & 361 & 315 \\ \hline
Mutation score      & 282/311 $\approx$  91\% & 239/311  $\approx$  77\%  & 361/402 $\approx$  90\%  & 315/402  $\approx$  78\%  \\ \hline
Average \# of killed mutants that could \textbf{not} be killed by the test cases selected by the other approach     & \cellcolor{yellow!25} 117 & \cellcolor{green!25} 128  & \cellcolor{yellow!25}193 & \cellcolor{green!25} 61  \\ \hline
\end{tabular}}
\vspace*{-.3cm}
\end{table}

In Table~\ref{tab:five models}, the number of selected test cases (i.e., the second row) is much lower compared to the number of mutants killed by these test cases (i.e., the third row). This difference, also observed in prior Simulink testing studies~\cite{matinnejad2018test,bartocci2023property,LiuLNBB16,GayRSWH16,MatinnejadNBBP13,LiuNLB19}, occurs because high block coverage for Simulink models can often be achieved using relatively few test cases~\cite{GayRSWH16}. Hence, a single test case can often kill multiple mutants. In addition, as shown in Table~\ref{tab:five models} and Figure~\ref{fig:test_cases_overlap}, transitioning from classical to requirements-aware mutation testing reduces the number of killed mutants and the number of selected test cases. This is because requirements-aware mutation killing is more challenging than classical mutation killing because mutants causing requirements violations are rarer than those causing output discrepancies.

\emph{Finding~1.}  \approach\  and FIM are complementary: not all the test cases identified by \approach\ could be identified by FIM and vice versa. Similarly, not all \approach\ mutants can be killed by the test cases selected by FIM and vice versa.

\emph{Finding~2.}  For classical mutation testing, FIM outperforms \approach. As shown in the cells highlighted yellow in Table~\ref{tab:five models}, a larger number of FIM mutants (193) cannot be killed by the test cases selected by \approach, compared to 117  \approach\ mutants that cannot be killed by test cases selected by FIM.  Under the classical mutation testing, FIM generates more mutants than \approach\ and its mutants, are more difficult to be killed. 

\emph{Finding~3.}  Under the requirements-aware notion of mutation testing, \approach\ outperforms FIM. Under this notion, test cases selected by FIM fail to kill twice the number of \approach\ mutants compared to the number of FIM mutants that the test cases selected by \approach\ fail to kill. Specifically, the cells highlighted green in Table~\ref{tab:five models} show that  a larger number of \approach\ mutants (128) cannot be killed by the test cases selected by FIM, compared to 61  FIM mutants that cannot be killed by test cases selected by \approach.  These results show that \approach\ generates mutants that are more representative of faults relevant to system requirements, and the faults that \approach\ mutants represent are harder to detect than those represented by FIM mutants.

\vspace*{-.15cm}
\section{Conclusion and Future Work}
\vspace*{-.15cm}
We presented a novel CodeBERT-based mutation testing approach for Simulink models. We show that our approach, using only  CodeBERT fine-tuning, can generate mutants that cover all known  mutation patterns related to individual Simulink blocks. Furthermore, we demonstrate that our approach complements -- and, under a requirements-aware notion of mutation testing, outperforms -- a state-of-the-art Simulink mutation testing tool. Our work  can be enhanced in several directions, such as assessing prompt-based mutation testing for Simulink and enhancing our approach to cover multi-block mutation patterns. Our replication package is available online~\cite{github}.

\section*{Acknowledgements}
We gratefully acknowledge funding from NSERC of Canada under the Discovery and Discovery
Accelerator programs.

\balance
\bibliographystyle{plain}
\bibliography{bibliography}

\end{document}